\documentstyle[12pt]{article}
\begin{document}
\begin{flushright}
KANAZAWA-99-15  \\ 
March, 2000
\end{flushright}
\vspace*{2cm}
\begin{center}
{\LARGE\bf Mass bound of the lightest neutral Higgs scalar 
in the extra U(1) models}\\
\vspace{1 cm}
{\Large  Y. Daikoku}
\footnote[1]{e-mail: daikoku@hep.s.kanazawa-u.ac.jp}
 ~{\Large and}~ {\Large D. Suematsu}
\footnote[2]{e-mail: suematsu@hep.s.kanazawa-u.ac.jp}
\vspace {1cm}\\
{\it Institute for Theoretical Physics, Faculty of Science,
Kanazawa University,\\
        Kanazawa 920-1192, Japan}
\end{center}
\vspace{2cm}
The upper mass bound of the lightest neutral Higgs scalar is studied 
in the $\mu$ problem solvable extra U(1) models by using the analysis of
the renormalization group equations.
In order to restrict the parameter space we take account of a 
condition of the radiative symmetry breaking and some phenomenological 
constraints. 
We compare the bound obtained based on this restricted parameter space
with the one of the next to the minimal supersymmetric standard model (NMSSM).
Features of the scalar potential and renormalization group 
equations of the Yukawa couplings among Higgs chiral supermultiplets 
are rather different between them. They can reflect in this bound.

\newpage
\noindent
{\large\bf 1.~Introduction} 

Low energy supersymmetry is one of the main subjects of present particle
physics. It is considered to solve a weak scale stability problem 
called the gauge hierarchy problem in the standard model (SM). 
Although we donot have any direct evidence for it,
it has been stressed that the gauge coupling unification 
occuring in a rather precise way in the minimal supersymmetric
extension (MSSM) of the SM may be an encouraging sign for the presence 
of the low energy supersymmetry.
In the MSSM its phenomenology crucially depends on the soft
supersymmetry breaking parameters and then it seems to be difficult
to make useful predictions unless we know how the supersymmetry breaks 
down.
However, there is an important exception that the lightest neutral 
Higgs scalar mass cannot be so heavy and it is mainly controled by the 
feature of the weak scale symmetry breaking \cite{higgs1}.
This is not heavily  dependent on the feature of the soft
supersymmetry breaking parameters at least at the tree level. 
Thus the knowledge of its possible upper bound is
crucial to judge the validity of the low energy supersymmetry from a 
viewpoint of the energy front of the accelerator experiment.
This aspect has been extensively studied taking account of a radiative 
correction mainly due to a large top Yukawa coupling \cite{higgs2}. 

It is well known that there still remains a hierarchy problem
called $\mu$ problem in the MSSM. Why a supersymmetric Higgsino
mixing term parametrized by $\mu$ is a weak scale cannot be explained
in the MSSM \cite{mu}.
A simple and promising candidate for its solution is an extension of
the MSSM by the introduction of an extra U(1) gauge symmetry and
a SM singlet field $S$ with a nonzero charge of this extra U(1) 
\cite{extra,extra1}.
The essential feature of this model is described by the following 
superpotential
\begin{equation}
W_{U(1)^\prime}=\lambda SH_1H_2 + kS\bar gg + h_tQH_2\bar T +\cdots,
\end{equation}
where $H_1$ and $H_2$ are usual doublet Higgs chiral superfields and 
the ellipses stand for the remaining terms in the 
MSSM superpotential other than the $\mu$
term and the top Yukawa coupling. 
In the second term $g$ and $\bar g$ stand for the extra color triplet
chiral superfields which are important to induce the $\mu$ scale.
In the superpotential 
$W_{U(1)^\prime}$ we also explicitly write the top 
Yukawa coupling because of its importance in the electroweak 
radiative symmetry breaking as in the case of the MSSM \cite{rad}.

The vacuum of these models is parametrized by the vacuum expectation 
values (VEVs) of Higgs scalar fields such as
\begin{equation}
\langle H_1\rangle=\left(\begin{array}{c}v_1 \\0\\ \end{array}\right),
\qquad \langle H_2\rangle=\left(\begin{array}{c}0 \\v_2\\ 
\end{array}\right),
\qquad
\langle S\rangle=u,
\end{equation}
where $v_1$ and $v_2$ are assumed to be positive
and $v_1^2+v_2^2=v^2(\equiv (174{\rm GeV})^2)$ should 
be satisfied\footnote{
In the following discussion we donot consider the spontaneous CP violation.
Under this assumption the sign of $u$ cannot be fixed freely
but it should be dynamically determined by finding the potential minimum.}. 
The vacuum in this model is parametrized by 
$\tan\beta=v_2/v_1$ and $u$.
The extra U(1) symmetry is assumed to be broken at the region not far
from the weak scale by a VEV of the scalar component of $S$ 
because of the radiative effect 
caused by the second term in $W_{U(1)^\prime}$ \cite{extra,extra1}
and then the $\mu$ scale is induced as $\mu=\lambda u$.
Thus in this model the sign of $\mu$ is fixed as the one of $u$ 
automatically.

This extra U(1) symmetry forbids a bare $\mu$ term 
in the superpotential and simultaneously makes the model free 
from the massless axion and tadpole problems. 
These features seem to make this model more promising than the next
to the minimal supersymmetric standard model (NMSSM) \cite{nmssm}
which is similar to this extra U(1) model but is extended only 
by a SM singlet chiral superfield 
$S$ with the superpotential 
\begin{equation}
W_{\rm NMSSM}=\lambda SH_1H_2 + {1 \over 3}\kappa S^3 + h_tQH_2\bar T +\cdots.
\end{equation}
It is also interesting that this kind of extra U(1) models can be often
obtained as the effective models of a lot of superstring models  
\cite{string}.
Various interesting features of this type of models have been studied in many
works by now [9$-$12].
Among the phenomenology of these models
the lightest neutral Higgs scalar mass is also an important target for 
the detailed investigation.
Of course, also in these models the lightest neutral Higgs scalar can be
expected to be generally not so heavy.
The interesting point is that its upper bound can be
calculable with no dependence on the soft supersymmetry breaking
parameters at least at the tree level as in the case of the NMSSM [13$-$17].
The dependence on the soft supersymmetry breaking parameters comes in
through the loop correction mainly due to the large top Yukawa
coupling and the second term of Eq. (1).  

In this paper we estimate the upper bound $m_{h^0}$ of this lightest 
neutral Higgs scalar mass on the correct vacuum.
The correct vacuum is determined as the radiatively induced minimum of 
the effective potential in the suitable parameter space.
In this approach we use the one-loop effective potential and solve 
the relevant renormalization group equations (RGEs) numerically.
We will pay our attention on the comparison of this upper 
bound with the one of the NMSSM within the phenomenologically 
allowable parameter region. 
In the NMSSM it has been known through many works that the triviality 
bound of a Yukawa coupling $\lambda$ of Higgs chiral superfields strictly 
control the upper bound of the lightest neutral Higgs mass 
\cite{nmssm2,nmssm3}.
Our approach is somehow different from this usual one.
We find the phenomenologically acceptable parameter subspace 
in the rather wide parameter space by
taking account of the radiative symmetry breaking condition and some 
phenomenological conditions such as the chargino mass and the charged 
Higgs scalar mass {\it etc.}.
The estimation of the upper mass bound of the lightest neutral Higgs
scalar is carried out in this restricted parameter subspace.
Although the result of this approach is necessarily dependent on the
assumption for the soft supersymmetry breaking parameters, we consider 
that it is possible to obtain the useful results by studying the 
wide region of the parameter space.
\vspace*{5mm}

\noindent
{\large\bf 2.~Extra U(1) models}

In this section we discuss more detailed features of the extra U(1) models
and give the basis of the present study.
Since the NMSSM is well known and discussed in many papers\cite{}, 
it is cinvenient to explain the points by using the extra U(1) models.
The superpotential of our considering extra U(1) 
models is defined by Eq. (1).
Soft supersymmetry breaking parameters are introduced as
\begin{eqnarray}
{\cal L}_{\rm soft}&=&-\sum_{i}m^2_{\phi_i}\vert\phi_i\vert^2
+\left(\sum_a{1\over 2}M_a\bar \lambda_a\lambda_a + {\rm h.c.}\right)
\nonumber \\
&+&\left(A_\lambda\lambda SH_1H_2 + A_kkS\bar gg + A_th_tQH_2\bar T +
{\rm h.c.}\right),
\end{eqnarray}
where the first two terms are mass terms of the scalar component
$\phi_i$ of each chiral supermultiplet and of gauginos $\lambda_a$.
We use the same notation for the scalar component as the one of the chiral
superfield to represent the trilinear scalar couplings in the 
last parentheses.
Other freedoms remaining in the models are extra matter contents 
and a type of extra U(1). 
On these points we confine our study into the typical extra U(1) models
derived from E$_6$, which are listed in Table 1.
At the TeV region they are assumed to have only one extra U(1) symmetry which
is broken only by the VEV of $S$ and give a solution to the $\mu$ problem
\cite{study2}.
As discussed in Ref. \cite{nmssm3} for the 
case of NMSSM, the extra matter contents affect indirectly the low 
energy value of the Yukawa coupling $\lambda$ through the influence 
on the running of the top Yukawa coupling. 
This is rather important to estimate the Higgs
mass bound. In the present model such kind of effects on the Yukawa 
couplings may also be expected but its effect is more complicated than 
the NMSSM as discussed later.
If we introduce the extra field contents arbitrarily,
the cancellation of the gauge anomaly may require to introduce 
the additional fields which again affect the running of Yukawa coupling 
$\lambda$ and so on.
Thus for the estimation of the Higgs mass bound it is important to fix the
matter contents in the anomaly free way in the present study.

\footnotesize
\begin{figure}[bt]
\begin{center}
\begin{tabular}{|c||ccccccccccc|}\hline
 & $Q$&  $\bar U$& $\bar D$&$L$  &$\bar E$&$H_1$&$H_2$&$g$&$\bar g$&$S$ &
$N$ \\\hline\hline
SM&{\small (3,2)}&{\small $(3^\ast,1)$}&{\small $(3^\ast,1)$}&{\small (1,2)}
&{\small (1,1)}&{\small (1,2)}&{\small (1,2)}&
{\small (3,1)}&{\small $(3^\ast,1)$}&{\small (1,1)}&{\small (1,1)}\\
$Y$ &$\displaystyle{1\over 3}$&$-\displaystyle{4\over
3}$&$\displaystyle{2\over 3}$&$-1$&2&$-1$&1&$-\displaystyle{2\over 3}$
&$\displaystyle{ 2\over 3}$&0&0 \\
$Q_\eta$&$-\displaystyle{ 2\over 3}$ &$-\displaystyle{ 2\over
3}$&$\displaystyle{ 1\over 3}$ &$\displaystyle{ 1\over
3}$&$-\displaystyle{ 2\over 3}$
&$\displaystyle{ 1\over 3}$&$\displaystyle{ 4\over 3}$&$\displaystyle{ 
4\over 3}$&$\displaystyle{ 1\over 3}$&$-\displaystyle{ 5\over 3}$
&$-\displaystyle{ 5\over 3}$ \\
$Q_{\xi_\pm}$&$\pm\displaystyle{ 1\over\sqrt6}$ &$\pm\displaystyle{
1\over\sqrt 6}$&$\pm\displaystyle{ 2\over\sqrt 6}$&
$\pm\displaystyle{ 2\over\sqrt 6}$&$\pm\displaystyle{ 1\over\sqrt
6}$&$\mp\displaystyle{ 3\over\sqrt 6}$&
$\mp\displaystyle{ 2\over\sqrt 6}$&$\mp\displaystyle{ 2\over\sqrt
6}$&$\mp\displaystyle{ 3\over\sqrt 6}$
&$\pm\displaystyle{ 5\over\sqrt 6}$& 0\\ \hline
\end{tabular}
\vspace*{-0.1cm}
\end{center}
{\footnotesize {\bf Table 1}\hspace{.5cm}
The charge assignment of extra U(1)s which are 
derived from $E_6$ \cite{mixing}.
These charges are normalized as $\displaystyle\sum_{i\in{\bf 27}}Q_i^2=20$.
}
\end{figure}
\normalsize

As the matter contents we assume the MSSM contents and additional 
extra matter fields 
$$
\Big[3( Q, \bar U, \bar D, L, \bar E )+( H_1, H_2 )\Big]_{\rm MSSM} +
3( g, \bar g )+2( H_1, H_2 )+3( S )+3( N ),
$$
which can be derived from three {\bf 27}s of E$_6$
shown in Table 1. This set satisfies the anomaly free conditions.
We can also add extra fields to these in 
the form of vector representations constructed from the fields listed
in Table 1.
Here we consider the following two cases as the additional extra chiral 
superfields
$$
{\rm (A)}\quad (H_a)+(H_a^\ast), \hspace*{1.5cm}
{\rm (B)}\quad (g+H_a+H_b)+(g^\ast+H_a^\ast+H_b^\ast),
$$   
where $a,b=1$ or 2 and 
the fields in the second parentheses come from {\bf 27}$^\ast$
of $E_6$. At least on the sector of SU(3)$_C \times$SU(2)$_L \times$U(1)$_Y$ 
these matter contents are the same as the one of 
[MSSM + $n({\bf 5}+{\bf 5}^\ast)$] where ${\bf 5}$ and ${\bf 5}^\ast$ are
the representations of the usual SU(5).
The case (A) corresponds to $n=3$ and (B) to $n=4$.
The $n=3$ is the critical value for the one-loop $\beta$-function of
SU(3). It makes this one-loop $\beta$-function be zero. 
The interesting point of these field contents is that 
the unification scale of SU(3)$_C \times$SU(2)$_L \times$U(1)$_Y$ is
not shifted from the MSSM one.
The $n=4$ case saturates the $\beta$-function for the 
pertubative running of gauge couplings up to the unification
scale $\sim 3\times 10^{16}$~GeV. 
Although this addition seems to be artificial, this type of spectrum
can be expected in the Wilson line breaking scenario of the E$_6$ 
type superstring model. 
We use these contents to compare the feature between the NMSSM and our
extra U(1) models\footnote{We should note that if every extra U(1) is broken 
near the unification scale, these models are equal to the NMSSM with
the equivalent extra matters as discussed in Ref. \cite{nmssm3}.}.  

The existence of multi-generation extra fields brings an ambiguity 
in Eq. (1).
The coupling $\lambda$ and $k$ can have generation indices for extra
fields such as $S$, $H_1$, $H_2$, $g$ and $\bar g$.
On this point we make the following assumption to make the argument 
simple.\footnote{
As far as we use Eq. (7) for the upper bound of the lightest neutral 
Higgs mass, this assumption seems to be reasonable.
This assumption affects the RGEs of some parameters and also one-loop 
correction to the Higgs scalar mass. Other cases will be discussed later.}
Only one $S$ can have the couplings in Eq. (1) and 
one pair of $(H_1, H_2)$ corresponding to the one of the MSSM alone
gets the VEVs.
Extra colored singlets $(g_i, \bar g_i)$ 
have a diagonal coupling to this $S$ as $k_i Sg_i\bar g_i$, where all 
the coupling constant $k_i$ show the same behavior in the RGEs because
$(g_i, \bar g_i)$ are completely symmetric for the generation index $i$
in the models. The fermion components of $g_i$ and $\bar g_i$ 
can get mass through this coupling.\footnote{
If we change the charge assignment for some fields, 
some extra fields can be heavy at the intermediate
scale as discussed in \cite{inter}. Although this kind of possibility
can be realized in the $\xi_{\pm}$ model, we donot consider it in this 
paper.}
On the other hand, the fermion components of the remaining 
$S$ which donot couple to the usual Higgsinos in $H_1$ and $H_2$ can get 
their masses through the one-loop correction.
From a viewpoint of the model construction, 
the serious phenomenological problem will be how the fermion
components in other remaining extra matter fields can get their
masses.
Although they can be generally massive through the gaugino mediated
one-loop diagrams, their magnitude seems not to be enough to satisfy
the phenomenological constraints.
In the $\xi_-$ model given in Table 1 we can introduce the 
intermediate scale through
the D-flat direction of $N$ and $N^\ast$ whose exsistence does not
affect our discussion in the later part of this paper.
If this is the case, they can have the weak scale masses through the
nonrenormalizable interactions in the superpotential such as $\displaystyle 
{1\over M_{\rm pl}}NN^\ast gg^\ast$.
Although this is phenomenologically important, it can be improved
by the suitable extension without changing the following results 
and thus we donot get involved in this point further here.

In our considering models the tree level scalar potential 
including the soft supersymmetry breaking terms can be written as
\begin{eqnarray}
V_0&=&{1\over 8}\left(g_2^2+g_1^2\right)\left(
\vert H_1\vert^2-\vert H_2\vert^2\right)^2 
+\left(\vert\lambda SH_1\vert^2+\vert\lambda SH_2\vert^2\right)\nonumber\\
&+&m_1^2\vert H_1\vert^2+m_2^2\vert H_2\vert^2 -(A_\lambda\lambda
SH_1H_2+{\rm h.c.})\nonumber\\
&+&{1\over 8}g_E^2\left(Q_1\vert H_1\vert^2
+Q_2\vert H_2\vert^2+Q_S\vert S\vert^2\right)^2 
+\lambda^2\vert H_1H_2\vert^2 +m_S^2\vert S\vert^2,
\end{eqnarray}
where $Q_1$, $Q_2$ and $Q_S$ are the extra U(1) charges of $H_1$, $H_2$ 
and $S$, respectively. 
The first two lines are found to have the corresponding 
terms in the MSSM if we
remind the fact that $\mu$ is realized as $\mu=\lambda u$.
The third line contains new ingredients. Its first term is a D-term 
contribution of the extra U(1) and
$g_E$ stands for its gauge coupling constant.

Potential minimum condition for Eq. (5) can be written as,
\begin{eqnarray}
\hspace*{-5mm}&&m_1^2=-{1\over 4}(g_2^2+g_1^2)(v_1^2-v_2^2)-{1\over
4}g_E^2Q_1(Q_1v_1^2+Q_2v_2^2+Q_Su^2) -\lambda^2(u^2+v_2^2)
+\lambda A_\lambda u{v_2\over v_1}, \nonumber \\
\hspace*{-5mm}&&m_2^2={1\over 4}(g_2^2+g_1^2)(v_1^2-v_2^2)-{1\over
4}g_E^2Q_2(Q_1v_1^2+Q_2v_2^2+Q_Su^2) -\lambda^2(u^2+v_1^2)
+\lambda A_\lambda u{v_1\over v_2}, \nonumber \\
\hspace*{-5mm}&&m_S^2=-{1\over 4}g_E^2Q_S(Q_1v_1^2+Q_2v_2^2+Q_Su^2)
-\lambda^2(v_1^2+v_2^2)+ \lambda A_\lambda{v_1v_2\over u}.
\end{eqnarray}
This constrains the soft SUSY breaking masses of Higgs scalars
around the weak scale. 
As the second derivative of $V_0$ in Eq. (5) we can derive the mass matrix of
the CP-even neutral Higgs scalar sector which is composed of three
neutral components $H_1^0,~H_2^0$ and $S$.
The goodness of this treatment has been discussed in the MSSM case
\cite{higgs2}
and we follow this argument.
If we note the fact that the smallest eigenvalue of any matrix 
is always smaller than any diagonal elements,  
we can obtain the tree level upper bound of this lightest Higgs scalar 
in an independent way of the soft supersymmetry breaking parameters 
by transforming the basis into the suitable one.
This upper bound can be written as \cite{bound,extra}
\begin{equation}
m_{h^0}^{(0)2} \le m_Z^2\left[\cos^22\beta +{2\lambda^2\over
g_1^2+g_2^2}\sin^22\beta+{g_E^2 \over
g_1^2+g_2^2}\left(Q_1\cos^2\beta+Q_2\sin^2\beta\right)^2\right], 
\end{equation}
where we used the potential minimization condition (6).
The first two terms correspond to the ones of the NMSSM in which
their behavior has been studied in many works \cite{hext,nmssm2,nmssm3}.
The running of a coupling constant $\lambda$ and its triviality bound
have been shown to be crucially dependent on the extra matters \cite{nmssm3}. 
The extra U(1) effect appears through the last term which is
its D-term contribution.
Equation (7) can show the different $\tan\beta$ dependence
from the one in the MSSM depending on the value of $\lambda$ 
and also the type of extra U(1).  
In the case of MSSM the upper bound of the lightest neutral Higgs mass
always increases with $\tan\beta$ in the region of $\tan\beta >1$.
If $\lambda ~{^<_\sim}~0.6$, the present model also shows the same
behavior.  
On the other hand, for the same region of $\tan\beta$ its upper bound can
decrease with increasing $\tan\beta$ when $\lambda ~{^>_\sim}~0.6$
which does not depend on the model so heavily. The NMSSM
shows the similar feature, which can be seen in
Ref. \cite{nmssm3}.
Although this may be potentially altered by the radiative correction,
it is one of the typical features coming from the $\lambda SH_1H_2$ 
in these models different from the MSSM.

We should note that the bound formula Eq. (7) is applicable only in the case 
of $u \gg v_1, v_2$ \footnote{
In the case of $u< v_1, v_2$ the diagonal element corresponding to $S$
can be smaller than the right-hand side of Eq. (7). In such a case we
cannot use Eq. (7) as the bound of the lightest neutral Higgs mass.
We will exclude it from our study.}.
In the extra U(1) model the value of $u$ can be constrained from below 
by the conditions on the mass of this extra U(1) gauge boson and 
its mixing with ordinary $Z^0$.
As far as we donot consider the special situation such as
$\tan^2\beta\sim Q_1/Q_2$ under which
the mixing with the ordinary $Z^0$ is negligible, the hierarchical condition 
$u > v_1, v_2$ should be imposed to satisfy the phenomenological
constraints on the extra $Z^\prime$ mass and its mixing with ordinary $Z^0$
\cite{mixing}. 
In the sufficiently large $u$ case $\lambda$ may 
be constrained into a limited range required by the successful 
radiative symmetry breaking at the weak scale so that $\lambda u(\equiv \mu)$ 
takes a suitable value. 
In the NMSSM this kind of constraint on $\lambda$ is expected to be
weaker than the one of the extra U(1) model since $u$ has no 
phenomenological constraint at this stage.
Anaway, we need the RGE study to check whether $\lambda$ can be
constrained in a substantial way by this condition.
\vspace*{5mm}

\noindent
{\large\bf 3.~The comparison of extra U(1) models and the NMSSM}

It is useful to discuss some qualitative features of the extra U(1)
models and the NMSSM in more detail 
before comparing the mass bound of the lightest
neutral Higgs in both models.
Although the extra U(1) models and the NMSSM have the similar 
feature related to
the $\mu$ term, they are expected to show rather different behavior in
the running of Yukawa couplings $k$, $\kappa$ and $\lambda$.
The top Yukawa coupling has the same one-loop RGE in both models as,
\begin{equation}
{d h_t \over d\ln\mu}={h_t\over 16\pi^2}\left(6h_t^2+\lambda^2-{16\over
3}g_3^2\right ).
\end{equation}
In the present field contents $g_3$ takes larger value at $M_X$ than
the one of the MSSM. Even if the initial value of $h_t$ takes the
large value like $O(1)$, the $\beta$-function in Eq. (8) can be small
due to the cancellation between a $h_t$ term and a $g_3$ term.
As a result, $h_t$ tends to stay near its initial value at the
intermediate scale independently whether it starts from a large value 
or a small value. This feature is shared by both models.
On the other hand, the one-loop RGEs of $\kappa$, $k$ and $\lambda$ 
are largely different from each other. They are witten as, in the NMSSM,
\begin{eqnarray}
&&{d \kappa \over d\ln\mu}={\kappa\over16\pi^2}
\left(6\kappa^2+6\lambda^2\right),
\nonumber \\
&&{d \lambda \over d\ln\mu}={\lambda\over 16\pi^2}
\left(3h_t^2+2\kappa^2+4\lambda^2\right ).
\end{eqnarray}
and in the extra U(1) model,
\begin{eqnarray}
&&{d k \over d\ln\mu}={k \over 16\pi^2}\left((3N_g+2)k^2
+2\lambda^2-{16\over 3}g_3^2\right ),
\nonumber \\
&&{d \lambda \over d\ln\mu}={\lambda\over 16\pi^2}
\left(3h_t^2+3N_gk^2+4\lambda^2\right ),
\end{eqnarray}
where $N_g$ is a number of the pair of the singlet colored fields $g$
and $\bar g$ which have a coupling to $S$.
In these RGEs we neglect the effect of 
gauge couplings $g_2$, $g_1$ and $g_E$ \footnote{
In these equations we cannot find the fixed ratio point other than
$k=0$ or $\lambda=0$ as far as $N_g\not=0$ even if we ignore $g_3$.
This is very different situation from the NMSSM which has been
discussed in Ref. \cite{bs}.}.
At first we consider the running behavior of $\kappa$ and $k$.
Since $k$ has an effect of $g_3$, it can be rather larger at the
intermediate scale than $\kappa$ which has no such effect and rapidly 
decreases according to lowering energy.  
This is important to determine the value of $u$ realized in both
models, which are mainly determined by $m_S^2$ at the low energy region.
They are controled by the one-loop RGE as
\begin{equation}
{d m_S^2 \over d\ln\mu}={1 \over 8\pi^2}\left(2\kappa^2(3m_S^2+A_k^2)
+2\lambda^2(m_S^2+m_{H_1}^2+m_{H_2}^2+A_\lambda^2)\right)
\end{equation}
in the NMSSM and 
\begin{equation}
{d m_S^2 \over d\ln\mu}={1\over 8\pi^2}\left(3N_gk^2(m_S^2+m_g^2
+m_{\bar g}^2+A_k^2)
+2\lambda^2(m_S^2+m_{H_1}^2+m_{H_2}^2+A_\lambda^2)\right),
\end{equation}
in the extra U(1) models.
The larger $k$ compared with $\kappa$ makes $m_S^2$ much more negative 
in the extra U(1) models. The larger value of $u$ is expected in the 
extra U(1) models if we remind Eq. (6). 

As easily seen from the RGE of $\lambda$ in the extra U(1) model,
the running of $\lambda$ is made fast by the existence of
the second term of Eq. (1) which is needed for the 
successful radiative symmetry breaking of these models \cite{extra}.
The one-loop $\beta$-function of the coupling $k$ has a
contribution of $g_3$ differently from the case of $\kappa$ in the
NMSSM. 
If we start $k$ and $\kappa$ from the large values at the
unification scale, this feature can keep $k$ rather large at the 
intermediate region
and then the running of $\lambda$ can be made fast by its effect 
compared with the one of $\kappa$ in the NMSSM.
This feature tends to make the value of $\lambda$ at the low energy
scale smaller compared with the NMSSM case if the same initial
value is adopted at least.
However, the initial value of $k$ and $\kappa$ should be controled from the 
requirement of the radiative symmetry breaking from our view point
since they play an important role in this phenomenon.  
We need the numerical analysis to study this aspect in more
 quantitative way.
The extra matter effects on the RGEs are also rather different
between the NMSSM and the extra U(1) models.
As far as all the couplings are
within the perturbative regime, the larger number of extra matter
fields make the gauge couplings at the unification scale larger. 
As pointed out in \cite{nmssm3}, 
in the NMSSM this indirectly makes the low
energy value of $\lambda$ larger through the smallness of $h_t$
 at the intermediate scale
whose $\beta$-function in Eq. (8) is kept small there.
On the other hand, in the extra U(1) models the runnings of $k$ and
$\lambda$ are simultaneously affected by the extra matters in both
direct and indirect manner, as is easily seen in Eq. (10). 

We know from these considerations that the resulting low 
energy values of $\lambda$ and
$u$ are rather different in both models. We should note that these
values affect the upper bound of the lightest neutral Higgs scalar
mass. Although Eq. (7) shows $\lambda$ is crucial to determine the tree 
level bound, $u$ is essential to determine the magnitude of the 
one-loop effect, especially in the extra U(1) models.
The radiative correction to Eq. (7) can be taken into account
based on the one-loop effective potential.
It is well-known that the one-loop contribution to the
effective potential can be written as \cite{effective,effective1}
\begin{equation}
V_1={1 \over 64\pi^2}{\rm Str}~ 
{\cal M}^4~ \left(\ln{{\cal M}^2 \over \Lambda^2}-{3\over 2}\right),
\end{equation}  
where ${\cal M}^2$ is a matrix of the squared mass of the fields
contributing to the one-loop correction and $\Lambda$ is a 
renormalization point.
In the usual estimation of the lightest neutral Higgs mass in the
NMSSM the top and stop contributions to $V_1$ are mainly considered as 
the relevant fields because of their large Yukawa coupling.
However, in the study of the extra U(1) models 
$k$ is rather large and then we should also take
account of the effect on ${\cal M}^2$ from the extra singelt colored  
chiral superfields $g$ and $\bar g$ which have a coupling with $S$.
A mass matrix of the stops is written as
\begin{equation}
\left(\begin{array}{cc}
\tilde m^2_{Q} +h_t^2v_2^2 & h_tv_2(-A_t + \lambda u\cot\beta)\\
h_tv_2(-A_t + \lambda u\cot\beta) & \tilde m^2_{\bar T} +h_t^2v_2^2 \\
\end{array}\right),
\end{equation}
and the one of the s-$g$quarks is expressed as
\begin{equation}
\left(\begin{array}{cc}
\tilde m^2_g +k^2u^2 & -A_kku + \lambda kv_1v_2\\
 -A_kku+ \lambda kv_1v_2 & \tilde m^2_{\bar g} +k^2u^2 \\
\end{array}\right), 
\end{equation}
where $\tilde m^2_{Q,\bar T},\tilde m^2_{g,\bar g}$ and $A_t, A_k$ are 
soft supersymmetry breaking parameters.
Here a D-term contribution is neglected as it has been done in many previous
investigations of the MSSM \cite{higgs2}.  
Mass eigenvalues of these mass matrices are respectively expressed as,
\begin{eqnarray}
&&\tilde m_{t_i}^2={1\over 2}(\tilde m^2_Q + \tilde m^2_{\bar T})
+h_t^2v_2^2 \pm \sqrt{{1 \over 4}(\tilde m^2_Q -\tilde m^2_{\bar T})
+h_t^2v_2^2(-A_t + \lambda u\cot\beta)^2}, \nonumber \\
&&\tilde m_{g_i}^2={1\over 2}(\tilde m^2_g + \tilde m^2_{\bar g})
+k^2u^2 \pm \sqrt{{1 \over 4}(\tilde m^2_g -\tilde m^2_{\bar g})
+(-A_kku + \lambda kv_1v_2)^2}.
\end{eqnarray}
If we estimate the upper bound of the lightest Higgs mass 
in the same procedure as the one used to obtain Eq. (7) by minimizing 
the one-loop effective potential $V_{\rm eff}=V_0+V_1$, the following 
one-loop correction should be added to the right-hand side of Eq. (7): 
\begin{equation}
\Delta m_{h^0}^2={1\over 2}\left({\partial^2 V_1 \over\partial
v_1^2}-{1\over v_1}{\partial V_1\over \partial v_1}\right)\cos^2\beta
+{1\over 2}{\partial^2 V_1 \over\partial v_1\partial v_2}\sin 2\beta
+{1\over 2}\left({\partial^2 V_1 \over\partial v_2^2}
-{1\over v_2}{\partial V_1\over \partial v_2}\right)\sin^2\beta.  
\end{equation}
From these we find that $u$ can crucially affect to the mass bound
through the one-loop effect of $g$quark sector in the extra U(1) models.
This addtional effect cannot be escapable as far as the occurence of the
radiative symmetry breaking is required. 
  
It may also be important to take account of the difference in both
models coming from some phenomenological constraints, in particular, 
the ones related to $\lambda$ and $u$.
Although this kind of constraints depend on the values of soft supersymmetry
breaking parameters, it may be useful to improve the upper bound
estimation based on the triviality bound of $\lambda$.
We should remind the fact that the chargino mass,
the charged Higgs mass and squark masses are dependent on 
$\lambda$ and $u$ \cite{study2}. 
The chargino and the charged Higgs scalar have the same
constituents as the MSSM.
However, they have a different mass formulas
from the MSSM \cite{mixing}. 
In both models the chargino mass is expressed as 
\begin{eqnarray}
&&m_{\chi^{\pm}}={1\over 2}\left(\lambda^2 u^2+2m_W^2+M_2^2\right)
\nonumber \\
&&\hspace*{1cm}\pm \sqrt{ {1\over 4}\left(2m_W^2\cos 2\beta+\lambda^2
u^2-M_2^2\right)^2+2m_W^2\left(-\lambda u\sin\beta +M_2\cos\beta\right)^2},
\end{eqnarray}
where $m_W$ and $M_2$ represent the W boson and the gaugino
$\lambda_2^{\pm}$ masses.
The charged Higgs scalar mass has the different mass formula between both 
models. In the extra U(1) models it is expressed as
\begin{equation}
m_{H^\pm}^2 =m_W^2\left(1-{2\lambda^2 \over g_2^2}\right)+{2A_\lambda
\lambda u\over \sin 2\beta},
\end{equation}
while in the NMSSM it is written as
\begin{equation}
m_{H^\pm}^2 =m_W^2\left(1-{2\lambda^2 \over g_2^2}\right)+{2(A_\lambda
\lambda u-\kappa\lambda u^2)\over \sin 2\beta}.
\end{equation}
Recently the lower bounds of these masses become larger and we may 
use these to put some constraints on $\lambda$ and $u$. 
Another important point to use Eq. (7) is that it must be smaller 
than other two diagonal mass matrix elements of the $3\times 3$ neutral Higgs
scalar mass matrix.
Especially the diagonal mass for the singlet Higgs scalar 
$S$ can give a substantial constraint on $u$. 
Its tree level formula is
\begin{equation}
m_{H^0_3}^2 ={1\over 2}g_E^2Q_S^2u^2+{A_\lambda\lambda v_1v_2\over u}
\end{equation}
in the extra U(1) models, while it is expressed as
\begin{equation}
m_{H^0_3}^2 =4\kappa^2u^2 +{A_\lambda\lambda v_1v_2\over
u}-A_\kappa\kappa u
\end{equation}
in the NMSSM.
This constraint may be substantial in the NMSSM
where there is no other clear constraint on the small $u$.
\vspace*{5mm}

\noindent
{\large\bf 4.~Numerical analysis and its results}

In this section we numerically estimate the bound of $m_{h^0}^2(\equiv 
m_{h^0}^{(0)2}+ \Delta m_{h^0}^2)$ by solving the RGEs and taking
account of the phenomenological constraints presented above.
In order to improve the one-loop effective potential \cite{loop}
We use two-loop RGEs for dimensionless coupling constants and one-loop 
ones for dimensional SUSY breaking parameters, for simplicity.
In this estimation we adopt the following procedure.
As the initial conditions for the SUSY breaking parameters we take
\begin{equation}
\tilde m_{\phi_i}^2=(\gamma_i\tilde m)^2, \qquad M_a=M, 
\qquad A_t=A_k=A_\kappa=A_\lambda =A,
\end{equation}
where $\tilde m^2$ is the universal soft scalar mass and 
we introduce the nonuniversality represented by $\gamma_i$ only among
soft scalar masses of $H_1, H_2$ and $S$. We comment on this point later.
These initial conditions are assumed to be applied at the scale where the 
coupling unification of SU(2)$_L$ and U(1)$_Y$ occurs. We donot require
the regolous coupling unification of SU(3)$_C$ but only impose the
realization of the low energy experimental value following Ref. \cite{nmssm3}.
For the extra U(1) coupling $g_E$ we use the same initial value as the one of 
U(1)$_Y$ at the unification scale $M_X$. 
The initial values of these parameters are surveyed through 
the following region,
\begin{eqnarray}
&&0\le h_t\le 1.2~~(0.1),\qquad 0\le k,~\vert\kappa\vert\le 2.0~~(0.2), 
\qquad 0\le \lambda \le 3.0~~(0.2), \nonumber\\
&&0\le M/M_S \le 0.8~~(0.2), \qquad
0\le \tilde m/M_S,~ |A|/M_S \le 3.0~~(0.3),
\end{eqnarray}
where in the parentheses we give the interval which we use in the
survey of these parameter regions.
Since the sign of $\kappa$ and $A$ affect the scalar potential, we
need to investigate both sign of them.
We also assume that the RGEs of the model are changed from the ones of the
supersymmetric extra U(1) models to the nonsupersymmetric ones 
at a supersymmetry 
breaking scale $M_S$ for which we take $M_S=1$~TeV as a typical numerical 
value \cite{hext,nmssm3}\footnote{In principle we should solve 
the RGEs of soft 
supersymmetry breaking parameters under the initial values given in
Eq. (15) in order to estimate this scale $M_S$. 
However, we donot take such a way here, for simplicity.
It is beyond the present scope to study the dependence of our 
results on the supersymmetry breaking scale $M_S$.}.

As a criterion for the choice of the correct vacuum,
we impose that the radiative symmetry breaking occurs correctly.
We check whether the potential minimum satisfying the conditions such as 
Eq. (6) improved by the one-loop effective potential can satisfy the
phenomenologically required conditions such as $v=174$~GeV and
$m_t=174$~GeV starting from the above mentioned initial conditions.
It is not so easy to find this solution under the completely universal
soft breaking parameters so that in our RGEs analysis 
we allow the nonuniversality
in the region $0.8 \le \gamma_i \le 1.2$ among soft supersymmetry breaking 
masses of Higgs scalars. 
The nonuniversality of soft scalar masses are generally expected in
the superstring models \cite{nonuniv}.
This treatment seems to be good enough 
for our purpose such as to estimate the upper mass bound of Higgs scalar.
We also additionally impose the following phenomenological conditions.\\
(i)~$m_{h^0}^{2}$ should be smaller than other
diagonal components of the Higgs mass matrix (see also footnote 3 and
the discussion related to Eqs. (13) and (14)).\\
(ii)~the experimental mass bounds on the charged Higgs bosons, 
charginos, stops, gluinos and $Z^\prime$ should be satisfied.
Here we require the following values:
\begin{eqnarray}
&&m_{H^\pm} \ge 67~{\rm GeV},\quad m_{\chi^\pm} \ge 72~{\rm GeV}, \quad
\tilde m_{t_{1,2}} > 67~{\rm GeV},\nonumber \\
&& M_3 \ge 173 ~{\rm GeV}, \quad m_{Z^\prime} \ge 500~{\rm  GeV}.
\end{eqnarray}
(iii)~the vacuum should be a color conserving one \cite{colorb}.\\
We adopt only the parameters set satisfying these criterions 
as the candidates of the correct vacua and calculate the Higgs mass bound 
$m_{h^0}^{2}$ for them. 

\input epsf 
\begin{figure}[tb]
\begin{center}
\epsfxsize=8cm
\leavevmode
\epsfbox{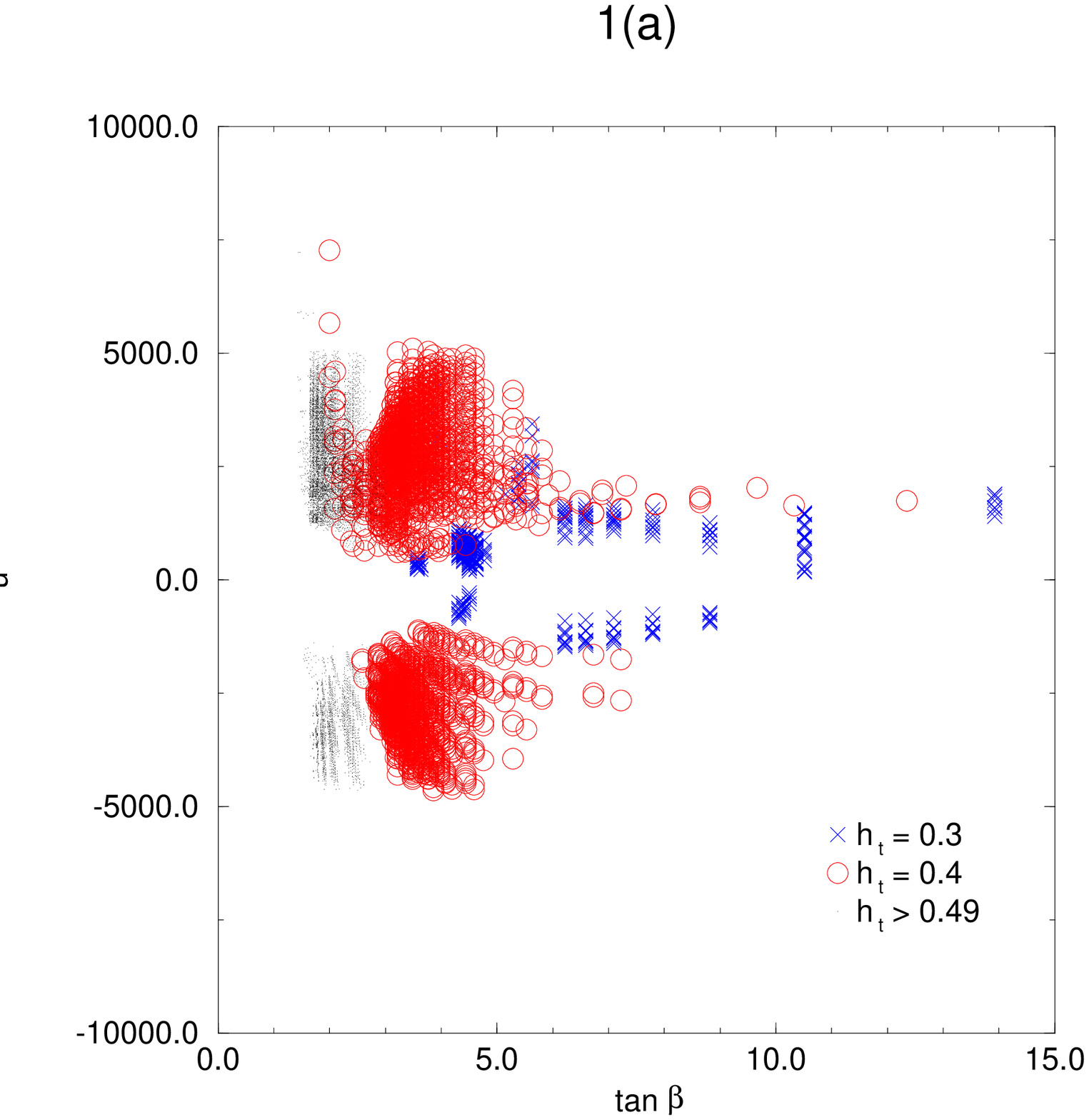}
\hspace*{0.4cm}
\epsfxsize=8cm
\leavevmode
\epsfbox{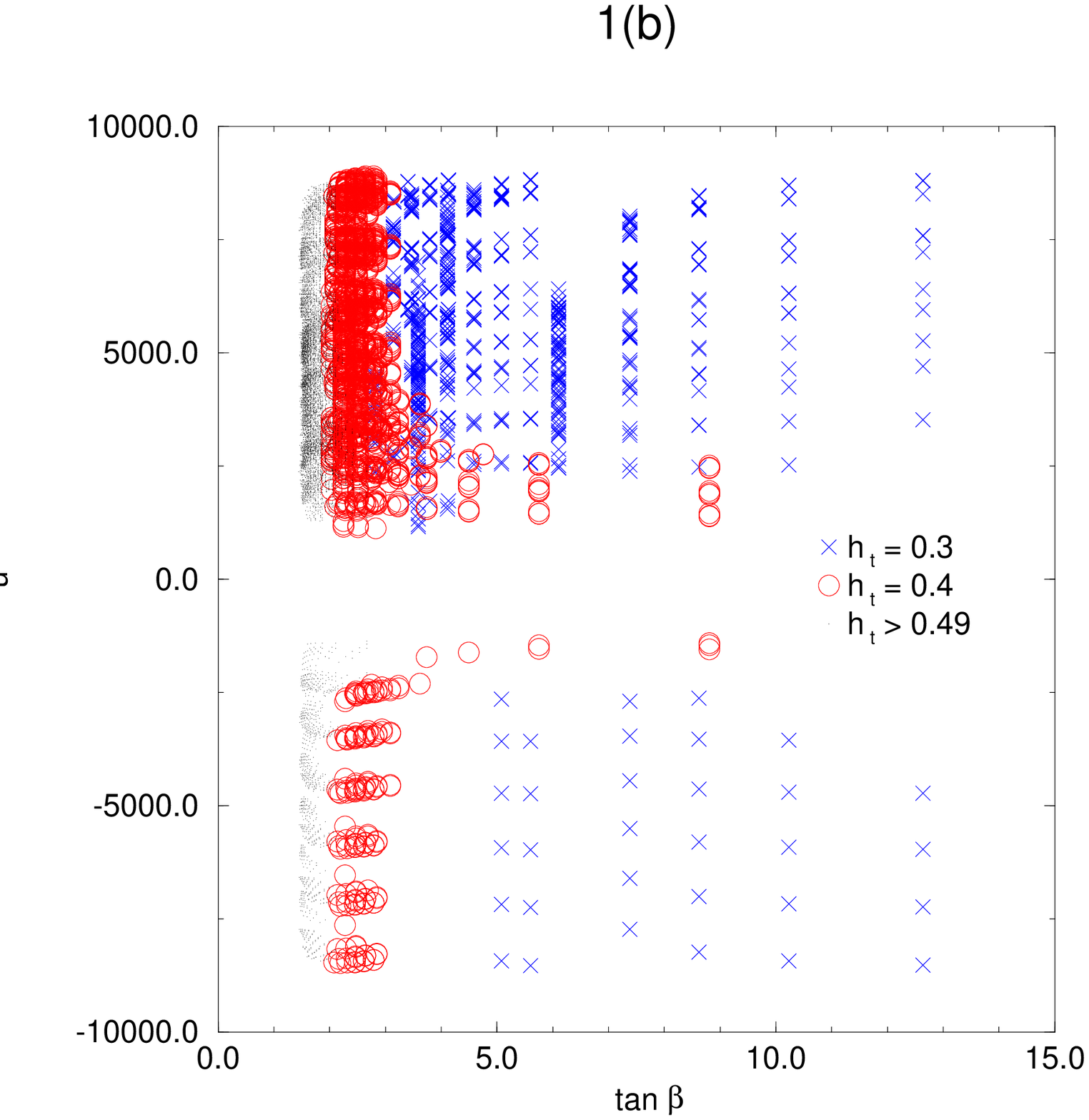}
\end{center}
\vspace*{-1cm}
{\footnotesize Fig.1~~\  Scatter plots of the radiative symmetry
breaking solutions in the ($\tan\beta$, $u$) plane for the NMSSM 1(a)
and the $\xi_-$ model 1(b). Solutions for the different $h_t(M_X)$
are classified.}
\end{figure}

At first in order to see the difference in the allowed vacuum between 
the NMSSM and the extra U(1) models we plot the radiative symmetry
breaking solutions for the present parameter settings 
in the $(\tan\beta, u)$ plane in Fig. 1.
Solutions are classified by the initial value of $h_t$ at $M_X$ 
into three classes which show rather different qualitative features.
As an example of the extra U(1) models we take the $\xi_-$ model here but
the $\eta$ model has been checked to show the similar feature to the $\xi_-$ model.
We take the case (A) as the extra matter contents.
Throught the present calculation an effect of the translation of 
the running mass to the pole mass \cite{run} is taken into account 
to determine $\tan\beta$.
We take $\tan\beta\le 15$ and neglect the large $\tan\beta$ solutions since
the bottom Yukawa coupling is assumed to be small in the RGEs so that 
in the present analysis the large $\tan\beta$ solutions cannot be 
recognized as the appropriate ones.  
Figure 1 shows that the $\xi_-$ model can have solutions in the larger
$u$ region of the $(\tan\beta, u)$ plane compared with the NMSSM.
As mentioned in the previous section, this is a result that $k$ can be
larger than $\kappa$ at the $m_t$ scale due to the SU(3) effect.
This is shown in Fig.2, where the values of $k(m_t)$, $\kappa(m_t)$ 
and $\lambda(m_t)$ corresponding to each solution are plotted for $\tan\beta$.
The soft scalar mass $m_S^2$ of the singlet Higgs scalar $S$ becomes 
much more negative in the
extra U(1) models than in the NMSSM.
In the sufficiently large $u$ region the potential minimum condition
for $u$ reduces to
\begin{equation}
u^2 =-{4 m_S^2\over g_E^2Q_S^2}\qquad {\rm
for~extra~U(1)}, \qquad
u^2=-{m_S^2 \over 2\kappa^2} \qquad {\rm for~NMSSM}. 
\end{equation}
In the NMSSM $u$ depends not only on $m_S^2$ but also on $\kappa$ and as
a result $u$ can take a rather large value.
In the $\xi_-$ model the smaller $u$ region such as 
$u~{^<_\sim}$~1~TeV is cut due to the experimental extra Z mass bound.
Also in the NMSSM very small $u$ seems to be forbidden.
This seems to be a result of the phenomenological conditions (i) and (ii). 

\begin{figure}[tb]
\begin{center}
\epsfxsize=8.4cm
\leavevmode
\epsfbox{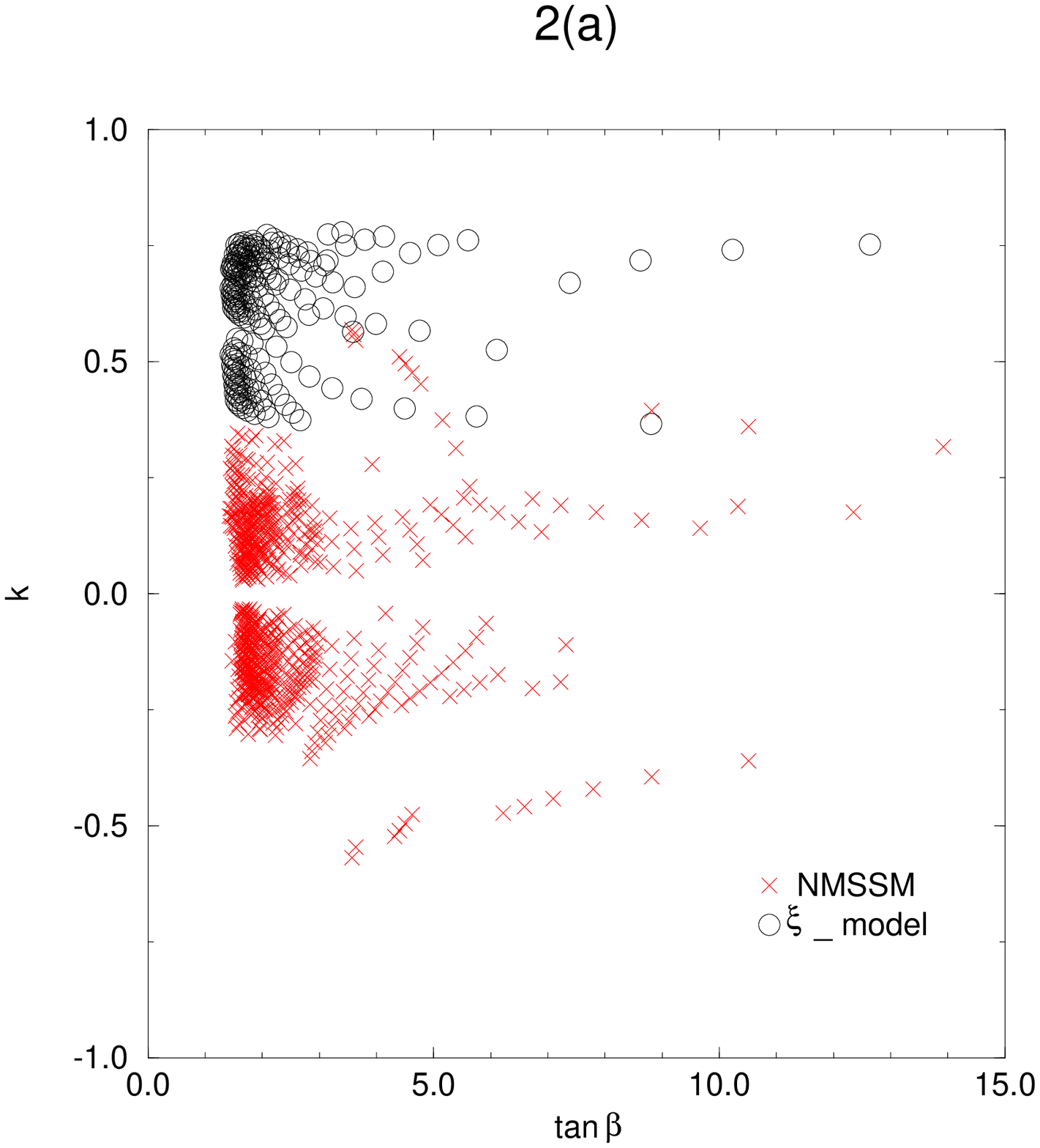}
\hspace*{-0.3cm}
\epsfxsize=8.4cm
\leavevmode
\epsfbox{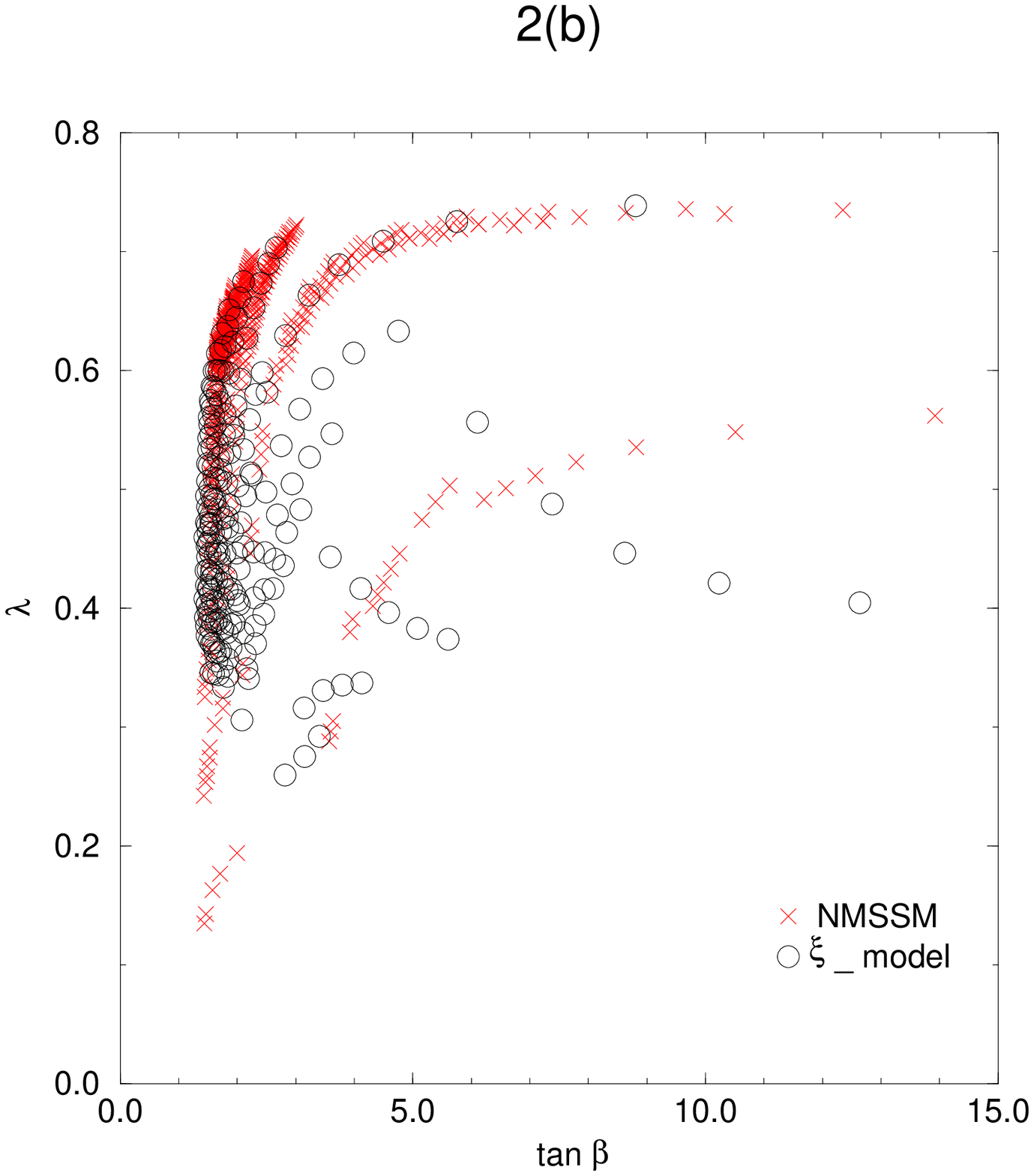}
\end{center}
\vspace*{-1cm}
{\footnotesize Fig.2~~\  Scatter plots of the radiative symmetry
breaking solutions for the NMSSM and the $\xi_-$ model in 
the $(\tan\beta, k~{\rm or}~\kappa)$ plane 2(a) and
the $(\tan\beta, \lambda)$ plane 2(b).  The values of $k$, $\kappa$
and $\lambda$ are the ones at $m_t$.}
\end{figure}

\begin{figure}[tb]
\begin{center}
\epsfxsize=8cm
\leavevmode
\epsfbox{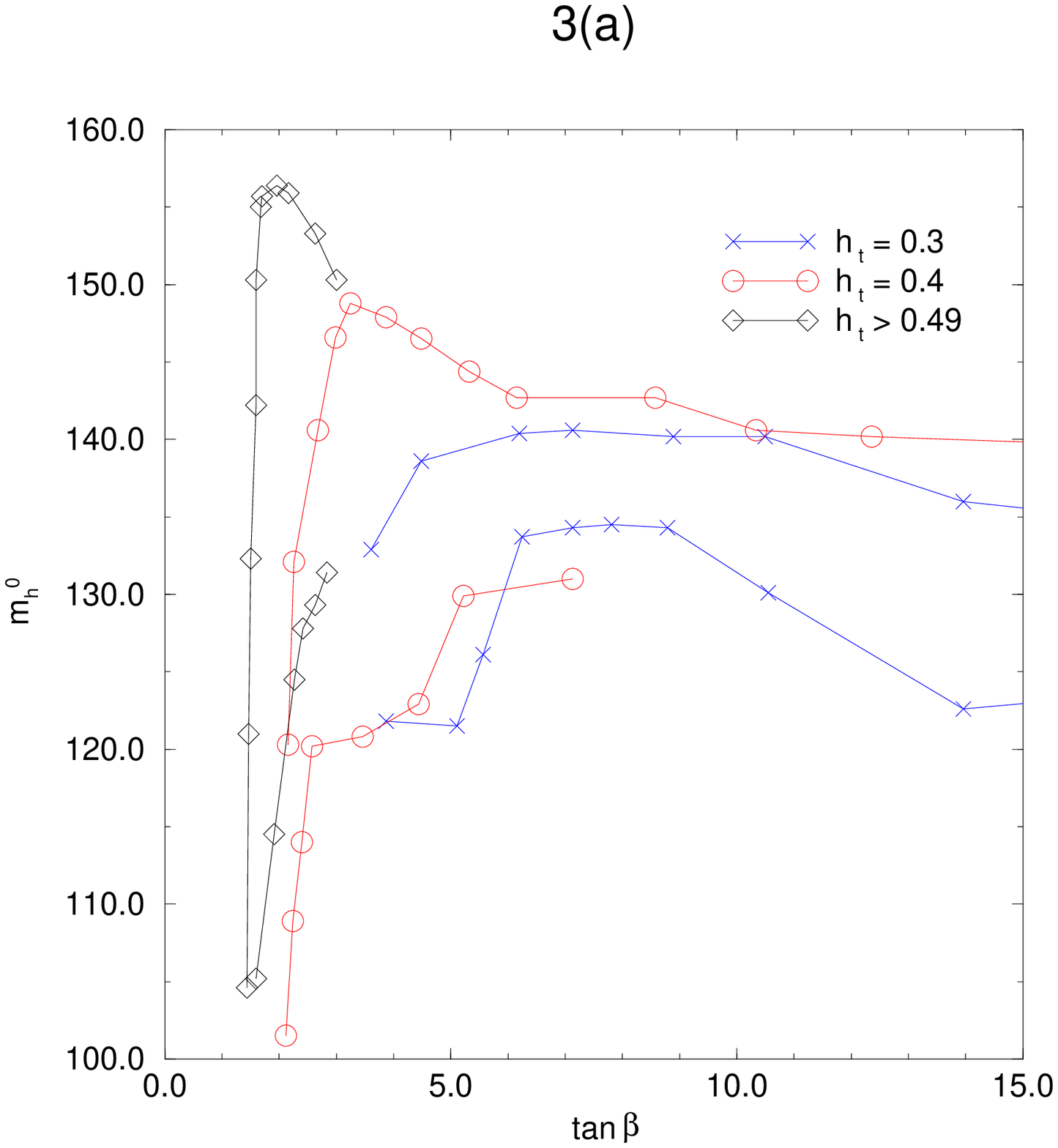}
\hspace*{-0.3cm}
\epsfxsize=8cm
\leavevmode
\epsfbox{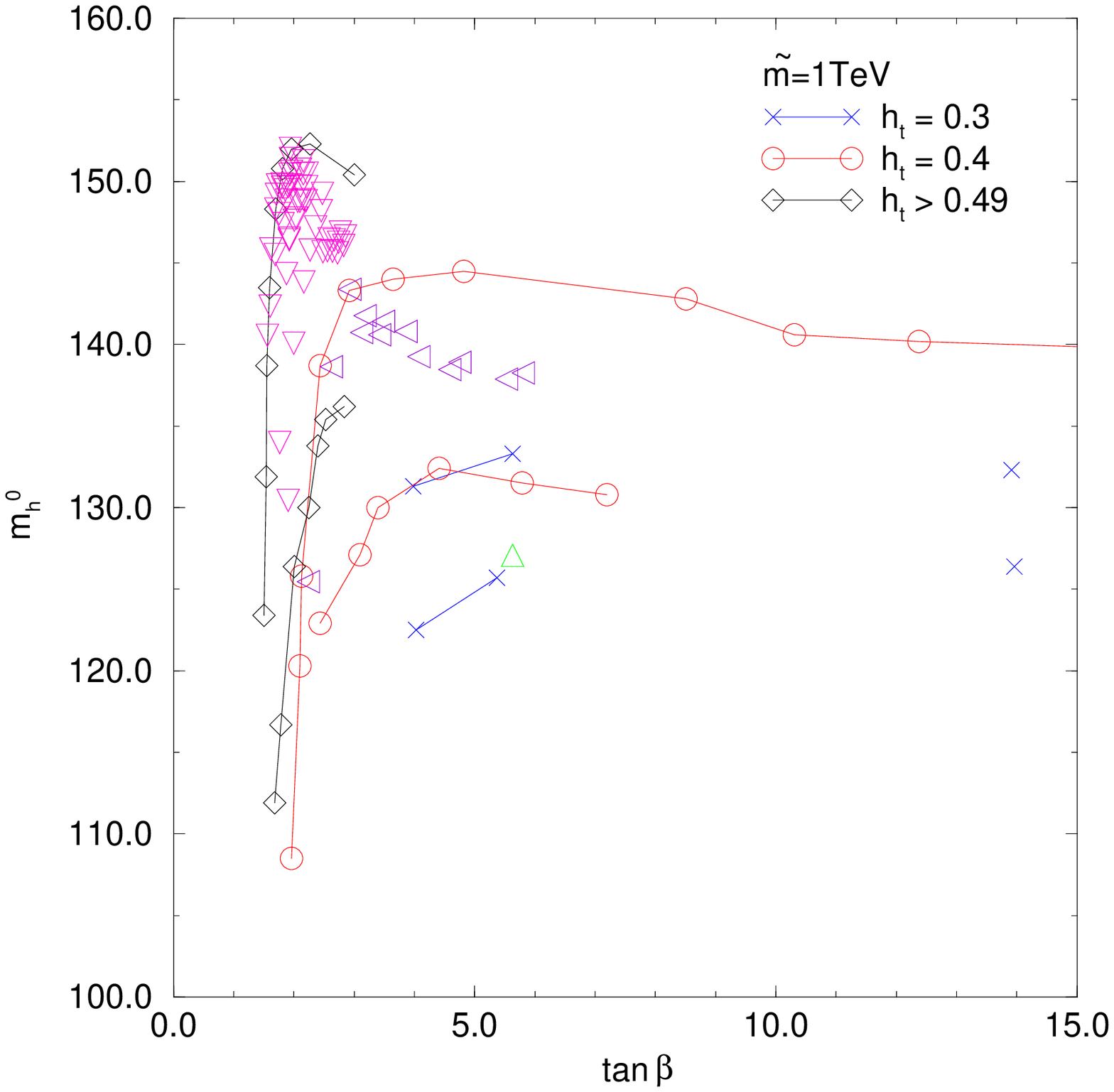}
\end{center}
\vspace*{-1cm}
{\footnotesize Fig.3~~\  Boundary values of $m_{h^0}$ of the lightest 
neutral Higgs mass as a function of $\tan\beta$ in the NMSSM. 
Full data are used to draw 3(a). In 3(b) we impose $\tilde
m(M_X)=1$~TeV. All solutions satisfying 2.4~TeV $\le u\le$ 2.6~TeV are
also plotted in 3(b).}
\end{figure}

The big qualitative difference of the vacuum in both models is 
that there can be large $u$ solutions for $\tan\beta~{^>_\sim}~5$ 
corresponding to $h_t(M_X)=0.3$ in the extra U(1) model.
One reason of this is that the smaller $\lambda(m_t)$ is
realized in the extra U(1) models than in the NMSSM.
This is clearly shown in Fig. 2(b). 
The discussion on this aspect has already given based on the RGE in
the previous section. 
On this point we should also note that in the $\tan\beta~{^>_\sim}~5$ region 
the small $\lambda(m_t)$ is allowed.
Thus $\mu=\lambda u$ can be in the suitable range even if $u$ is large.
However, the boundary value of $u$ seems not to 
have so strong dependence on $\lambda(m_t)$ in both models and the value 
of $\lambda u$ does not seem to be strictly restricted by the
radiative symmetry breaking at least within the parameter region
searched in this paper. 
 
\begin{figure}[tb]
\begin{center}
\epsfxsize=8cm
\leavevmode
\epsfbox{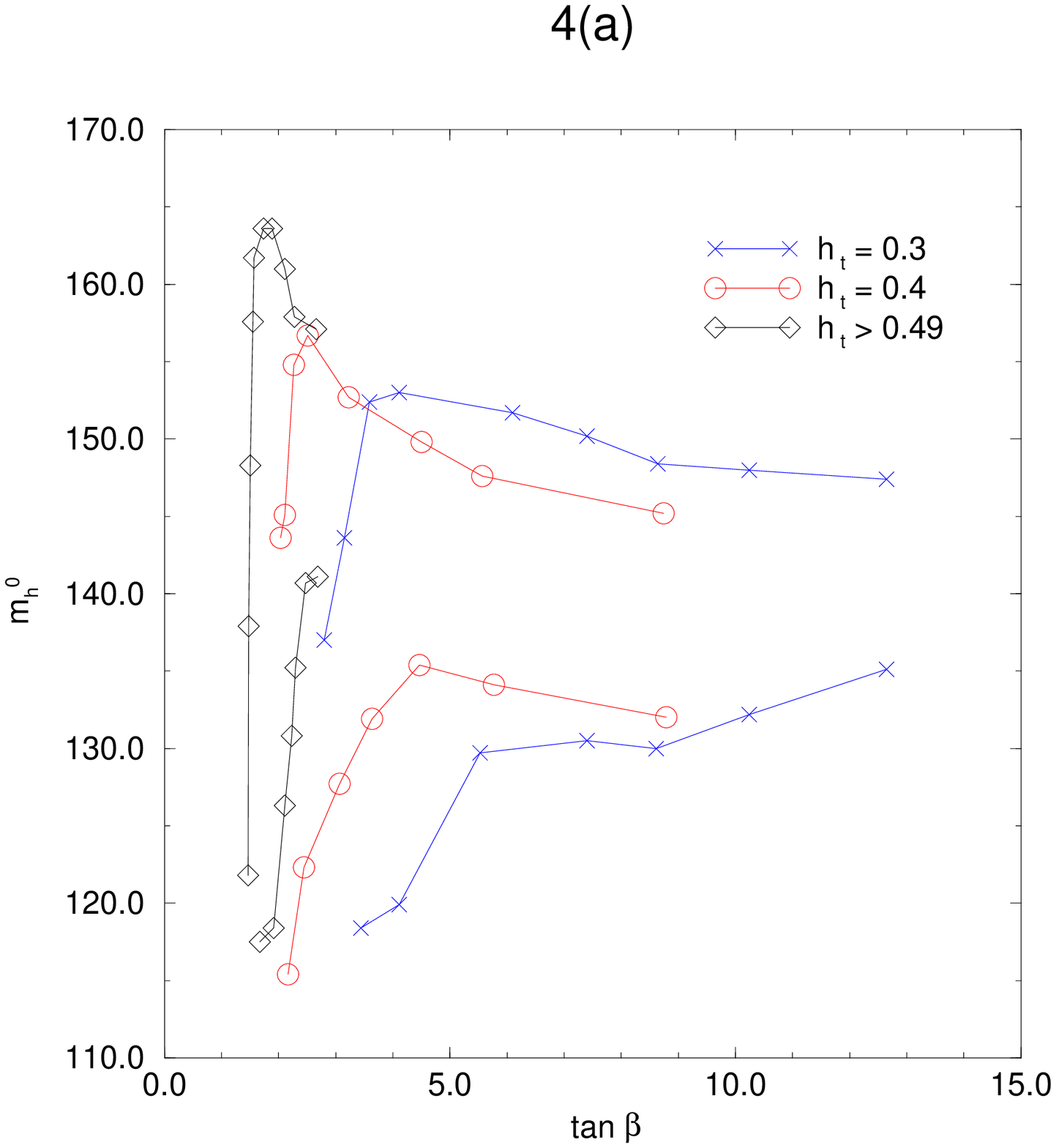}
\hspace*{-0.3cm}
\epsfxsize=8cm
\leavevmode
\epsfbox{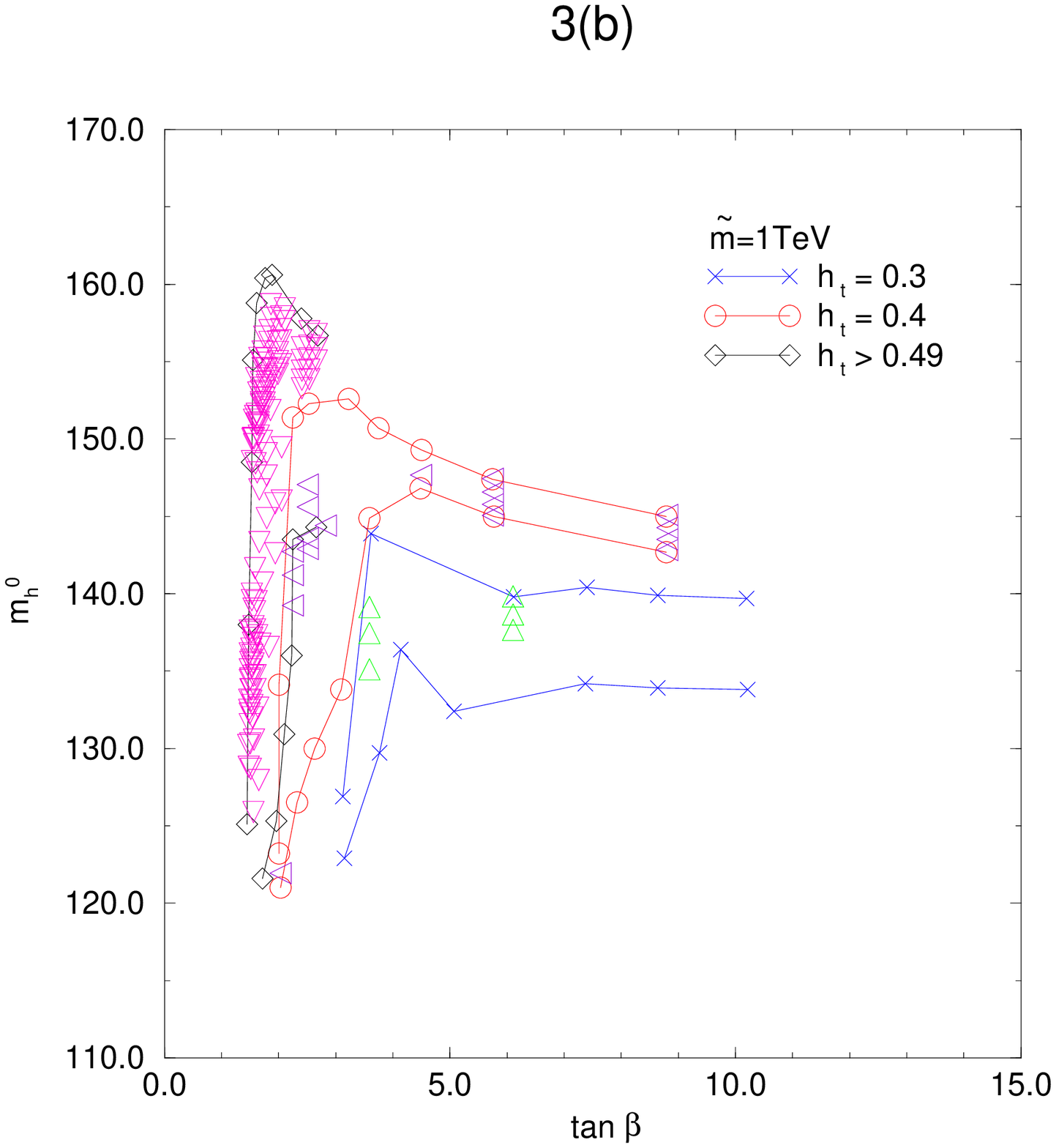}
\end{center}
\vspace*{-1cm}
{\footnotesize Fig.4~~\ Boundary values of $m_{h^0}$ of the lightest 
neutral Higgs mass as a function of $\tan\beta$ in the $\xi_-$ model. 
Full data are used to draw 4(a). In 4(b) we impose $\tilde
m(M_X)=1$~TeV. All solutions satisfying 2.4~TeV $\le u\le$ 2.6~TeV are
also plotted in 4(b).}
\end{figure}

\begin{figure}[tb]
\begin{center}
\epsfxsize=8cm
\leavevmode
\epsfbox{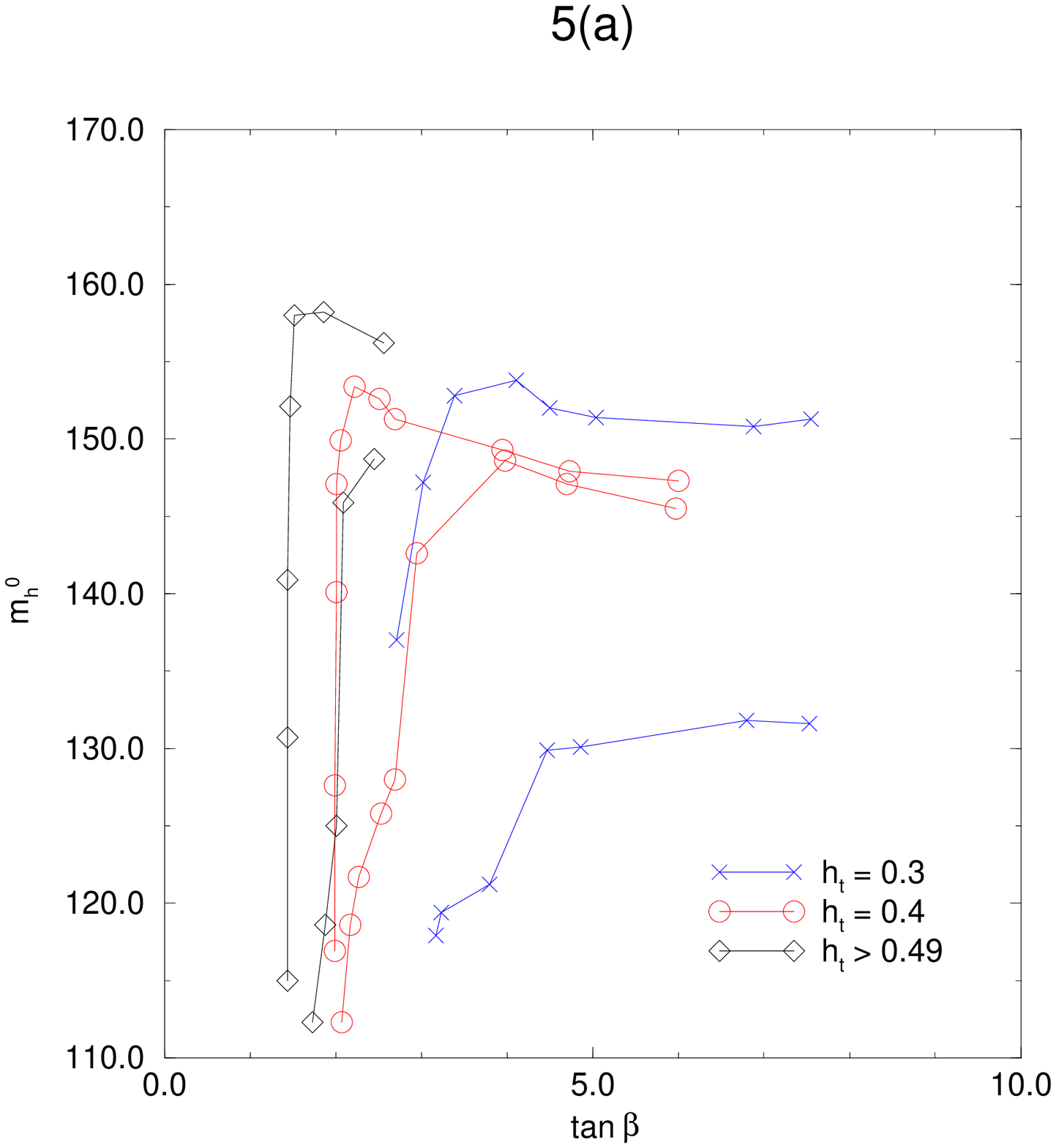}
\hspace*{-0.3cm}
\epsfxsize=8cm
\leavevmode
\epsfbox{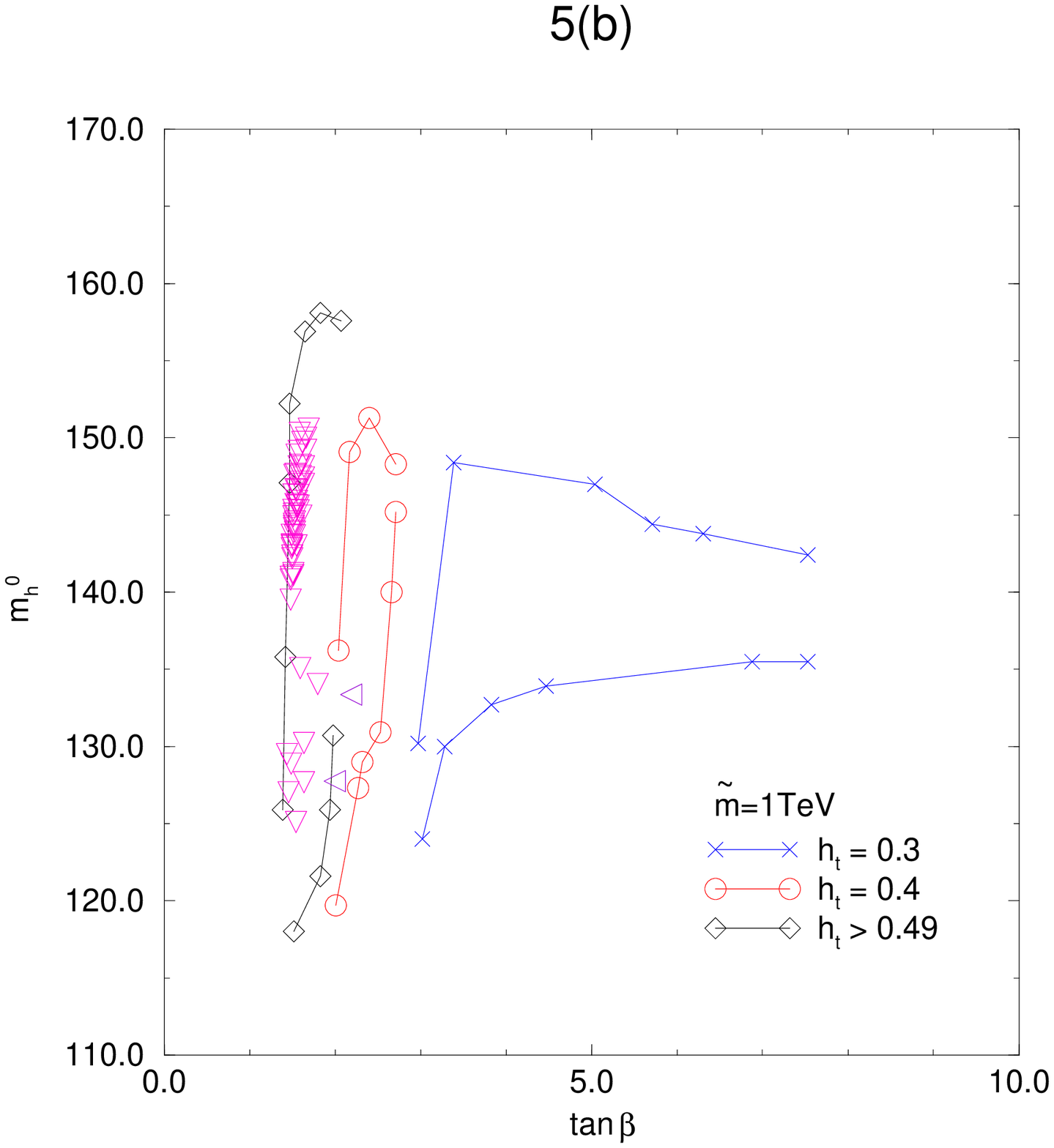}
\end{center}
\vspace*{-1cm}
{\footnotesize Fig.5~~\ Boundary values of $m_{h^0}$ of the lightest 
neutral Higgs mass as a function of $\tan\beta$ in the $\eta$ model. 
Full data are used to draw 5(a). In 5(b) we impose $\tilde
m(M_X)=1$~TeV. All solutions satisfying 2.4~TeV $\le u\le$ 2.6~TeV are
also plotted in 5(b).}
\end{figure}

\begin{figure}[tb]
\begin{center}
\epsfxsize=8cm
\leavevmode
\epsfbox{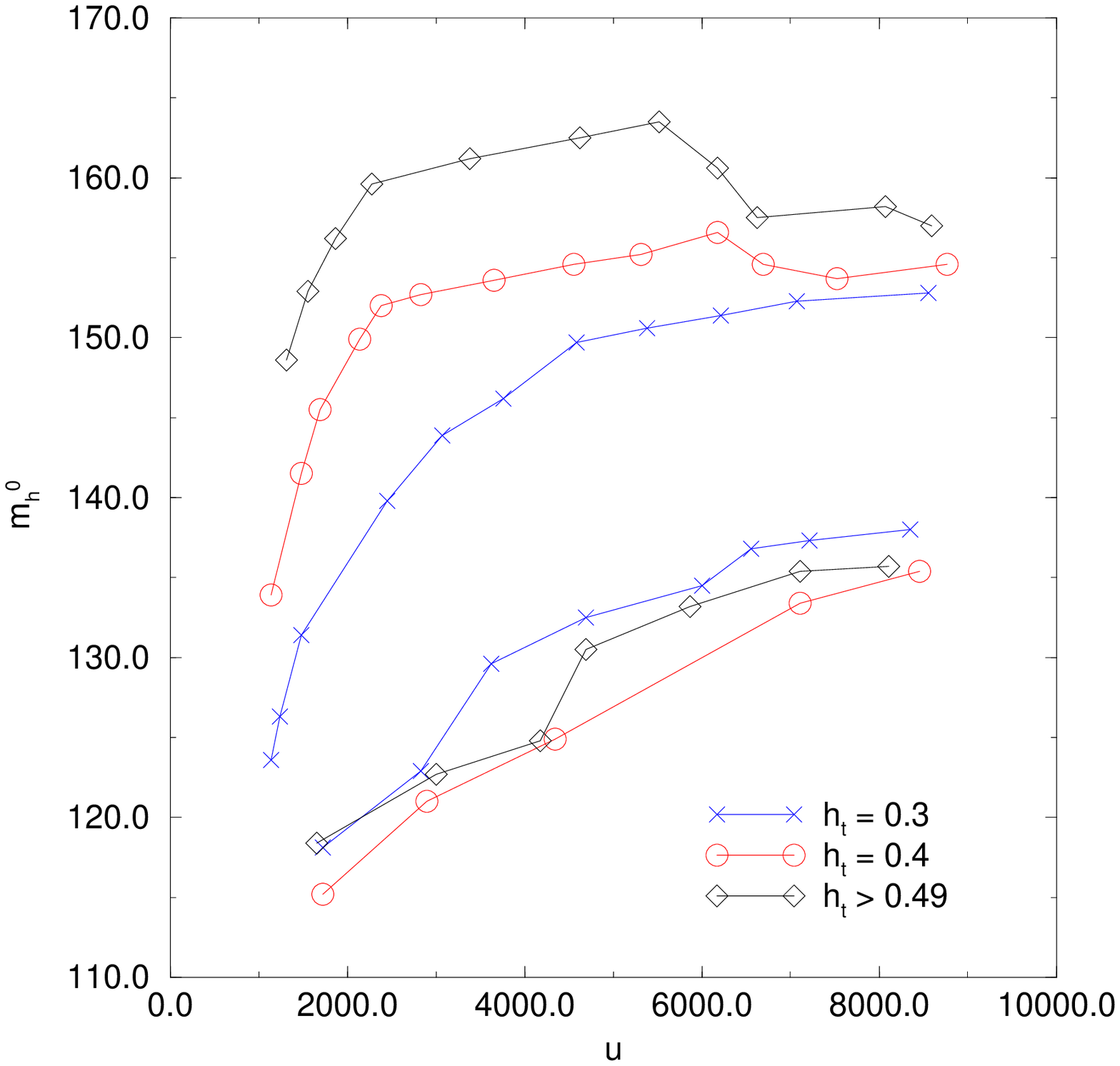}
\end{center}
\vspace*{-1cm}
{\footnotesize Fig.6~~\ Boundary values of $m_{h^0}$ of the lightest 
neutral Higgs mass as a function of $u$ in the $\xi_-$ model.}
\end{figure}

In Figs. 3$\sim$5 we give the results of our numerical estimations 
of $m_h^0$ for each model.
In these figures we plot the boundary values of $m_h^0$ for the
parameters obtained as the solutions of our radiative symmetry breaking study.
In each figure (a) the upper and lower boundaries of $m_h^0$ are drawn 
by using the all solutions obtained under the intial values shown in (24).
In oder to show the $h_t(M_X)$ dependence of $m_h^0$ we classify the
solutions into three classes and draw them separately.
In figures (b) we plot the upper and lower boundaries of $m_h^0$ 
for the remaining solutions after imposing the additional condition 
$\tilde m=1$~TeV. We also add the scatter plots of the all solutions
corresponding to 2.4~TeV $< u<$ 2.6~TeV in the same figures.
They are represented by three kinds of triangles corresponding 
to each $h_t(M_X)$. 
As a common feature in all models, we find that the larger 
$h_t(M_X)$ realizes the smaller $\tan\beta$ and then brings the larger 
contribution of the second term of Eq. (7). 
Thus the largest $\lambda(m_t)$ in the small $\tan\beta$ in Fig. 2(b) 
gives the largest $m_h^0$. Although $\lambda(m_t)$ in the extra U(1)
models can be smaller than the one of the NMSSM as shown in Fig. 2(b), 
the boundary values of 
$m_h^0$ is larger in the extra U(1) models than in the NMSSM by a few to ten 
GeV.
This is mainly due to the extra contribution to Eq. (17) coming from the
singlet colored fields $(g_i, \bar g_i)$. Since the existence of this
contribution is the basic feature of the present extra U(1) models,  
the boundary value of $m_h^0$ is generally expected to be larger than 
the one of the NMSSM inspite of the running feature of the Yukawa
coupling $\lambda$. 
This one-loop effect is large enough to cancel
the difference of $\lambda(m_t)$ in the second term of Eq. (7).
In our studying parameters space the largest value of $m_h^0$ is
\begin{equation} 
m_h^0~{^<_\sim}~ 156~{\rm GeV} {\rm (NMSSM)}, \quad
m_h^0~{^<_\sim}~ 164~{\rm GeV} {\rm (NMSSM)}, \quad
m_h^0~{^<_\sim}~ 158~{\rm GeV} {\rm (NMSSM)}.
\end{equation}
By comparing (a) and (b) in Figs. 3$\sim$5 we can get the tendency how
the solutions are restricted when we reduce the parameter space.
The change of $\tilde m$ and $u$ mainly affect the one loop
contribution through the mass matrices (14) and (15).

In Fig. 6 we plot the boundary value of $m_h^0$ for $u$ in the $\xi_-$ 
model. This shows the tendency that the larger $u$ gives the larger
value of $m_h^0$. This is expected from the one-loop contribution of
the extra singlet colored fields $(g_i, \bar g_i)$. 
From this figure we can read off the relation between $m_{Z^\prime}$ 
and $m_h^0$ by using $m_{Z^\prime}^2 \sim g_E^2Q_S^2u^2/2$.
The lower bound of $m_{Z^\prime}$ in Fig. 6 is about 600~GeV
where we used $g_E(m_t)=0.36$.
The conditions (i) and (ii) also determine the lower bound of $u$ in
the extra U(1) models.

Finally we give a few comments on some points related to the extra matters.
We also studied the case (B) of the extra matter contents for the same 
parameter settings as the above study.
In that case, as a common feature we can find, it becomes rather difficult to 
satisfy both of the radiative symmetry breaking conditions and the 
phenomenological conditions (i) to (iii) compared with the case (A).
The number of solutions in the case (B) is drastically 
less than in the case (B). 
Since the value of $g_3(M_X)$ increases, $h_t(m_t)$ and $k(m_t)$
becomes larger. In fact, the initial value of $h_t$ in the wide region 
such as $0.2\le h_t(m_t)\le 0.9$ results in only the small $\tan\beta$
(larger $h_t(m_t)$) solution such as $\tan\beta~{^<_\sim}~1.8$.
This also makes $\lambda(m_t)$ smaller.
The larger $\tan\beta$ solutions disappear and the value of $\vert
u\vert$ is shifted upward.
However, the upper boundary value of $m_h^0$ behaves in the different
way between the NMSSM and the extra U(1) models.
Although in both models $m_{h^0}$ becomes smaller in the region of 
$\tan\beta~{>_\sim}~2$, the behavior is different at $\tan\beta~{<_\sim}~2$. 
In the NMSSM it is a little bit larger than the one of case (A).
On the other hand, it becomes smaller than the one of case (A) 
by a several GeV in the extra U(1) models.
Here we should remind the fact that even if $\lambda(m_t)$ is smaller 
$m_{h^0}$ can be larger in the case that corresponding $tan\beta$ is
smaller. The difference in the RGE of $\lambda$ in both model
is also important in this behavior.
To have more confident quantitative results in this case we need to 
search the parameter space in the finer way.
We also changed the number of $(g_i, \bar g_i)$ which couples to $S$
in the superpotential (1) in the case (A). If we decrease this number from
three to one, the boundary values of the allowed $m_h^0$ become
larger. This reason is considered as follows. 
Although this decrease reduces the number of fields contributing to
the one-loop effective potential,
this also decreases the $N_g$ value in Eq. (10). As a result the larger $k$ and
$\lambda$ are realized at the low energy region. The larger $k$ also
brings the larger $u$. 
The contribution to the one-loop effect per a field can be larger.
Thus the decrease of the number of 
$(g_i, \bar g_i)$ which couples to $S$ causes the increase of $m_h^0$
at not only the tree level but also the one-loop level. 
\vspace*{5mm}

\noindent
{\large\bf 5.~Summary}

There are two well-known low energy candidates to solve the $\mu$ problem
in the MSSM. These are the NMSSM and the extra U(1) models.
We have estimated the upper bound of the lightest neutral Higgs mass 
in both models.
Apart from a Higgs coupling $\lambda SH_1H_2$,
there is a typical coupling $\kappa S^3$ in the NMSSM
and $k S g \bar g$ in the extra U(1) models.
In the NMSSM $\kappa$ plays a crucial role in the evolution of
$\lambda$ which dominantly determines the tree level mass 
bound of the lightest neutral Higgs scalar and in the radiative 
symmetry breaking.
In the extra U(1) models the introduction of the extra colored fields 
$g, \bar g$  and its coupling with the singlet Higgs $S$ are
crucial to cause the radiative symmetry breaking at the weak scale
successfully. This coupling can also affect the running of the
coupling constant $\lambda$.
We focussed our attention on these points and estimated the the upper
bound of the lightest neutral Higgs mass in both models. 
In this estimation we additionally imposed some phenomenological 
constraints related to $\lambda$ and the VEV of $S$
coming from, for example, the mass bounds of the 
charginos, the charged Higgs scalars and the $Z^\prime$ boson.
We solved the minimum conditions of the one-loop effective
potential improved by the RGEs for the couplings and soft
supersymmetry breaking parameters whose initial conditions are taken 
in the suitable region.
We estimated the upper bound of the lightest neutral Higgs scalar 
for the parameters which bring the phenomenologically correct
potential minimum.
Its tree level contribution due to $\lambda$ can be smaller 
in the extra U(1) models than in the NMSSM.
However, there is the extra one-loop contribution originated from
the Yukawa coupling $kSg\bar g$ and this makes its upper bound
larger in the extra U(1) than in the NMSSM by a few to ten GeV. 
It is interesting enough that the upper bound of the lightest 
neutral Higgs scalar in the extra U(1) models is not so different
from the one of the NMSSM.
The extra U(1) models may be an equal candidate to the NMSSM 
for the experimental Higgs search.

\vspace*{1cm}
This work has been partly supported by the
a Grant-in-Aid for Scientific Research from the Ministry of Education, 
Science and Culture(\#11640267 and \#11127206).

\end{document}